\documentclass[prl,superscriptaddress,a4paper,aps,twocolumn,showpacs]{revtex4}
\usepackage{graphicx,bm,color}

\def\--{\negthinspace - \negthinspace}

\begin{document}

\title{
  Computational spectroscopy of helium--solvated molecules: \\
  effective inertia, from small He clusters toward the nano-droplet
  regime 
}

\author{
  Stefano Paolini
}
\email{
  paolini@sissa.it
}
\affiliation{
  SISSA -- Scuola Internazionale Superiore di Studi Avanzati,
  via Beirut 2-4, 34014 Trieste, Italy
}
\affiliation{
  INFM--DEMOCRITOS National Simulation Center, Trieste, Italy
}

\author{
  Stefano Fantoni
}
\email{
  fantoni@sissa.it
}
\affiliation{
  SISSA -- Scuola Internazionale Superiore di Studi Avanzati,
  via Beirut 2-4, 34014 Trieste, Italy
}
\affiliation{
  INFM--DEMOCRITOS National Simulation Center, Trieste, Italy
}

\author{
  Saverio Moroni
}
\email{
  moroni@caspur.it
}
\affiliation{
  INFM--DEMOCRITOS National Simulation Center, Trieste, Italy
}

\author{
  Stefano Baroni
}
\email{
  baroni@sissa.it
}
\affiliation{
  SISSA -- Scuola Internazionale Superiore di Studi Avanzati,
  via Beirut 2-4, 34014 Trieste, Italy
}
\affiliation{
  INFM--DEMOCRITOS National Simulation Center, Trieste, Italy
}

\date{
  \today
}

\begin{abstract}
  Accurate computer simulations of the rotational dynamics of linear
  molecules solvated in He clusters indicate that the large-size
  (nano-droplet) regime is attained quickly for light rotors (HCN, CO)
  and slowly for heavy ones (OCS, N$_2$O, CO$_2$), thus challenging
  previously reported results. Those results spurred the view that the
  different behavior of light rotors with respect to heavy
  ones---including a smaller reduction of inertia upon solvation of
  the former---would result from the lack of {\em adiabatic following}
  of the He density upon molecular rotation. We have performed
  computer experiments in which the rotational dynamics of OCS and HCN
  molecules was simulated using a fictitious inertia appropriate to
  the {\em other} molecule. These experiments indicate that the
  approach to the nano-droplet regime, as well as the reduction of the
  molecular inertia upon solvation, is determined by the anistropy of
  the potential, more than by the molecular weight. Our findings are
  in agreement with recent infrared and/or microwave experimental data
  which, however, are not yet totally conclusive by themselves.
\end{abstract}

                                                                                                             

\maketitle

\section{Introduction}

The recent discovery that small molecules solvated in He clusters
display a sharp rotational spectrum with well resolved lines is a
spectacular manifestation of He superfluidity in such extreme confined
conditions \cite{toennies-OCS,jaeger-OCS}. Recent developments in
quantum many-body simulation techniques are allowing for the
determination of the first few excited states of interacting boson
systems \cite{NoAntri-Cornell,NoAntri-PRL,POITSE}, as well as for an
understanding of the relations existing between structure, dynamics,
and superfluidity in these systems \cite{NoAntri-2003,whaley-2003}. In
particular, it has now become possible to {\em predict} the dependence
of the effective molecular rotational constant, $B$, on the cluster
size, $N$, for clusters up to a few tens of He atoms \cite{Genova}.

Results based on the so-called {\em projection operator imaginary-time
  spectral evolution} (POITSE, an implementation of diffusion quantum
Monte Carlo aimed at estimating excitation energies) provide the
following scenario for the dependence of $B$ upon $N$
\cite{whaley-2003,focus_article,toennies}: {\em i)} the inertia of
heavy rotors is reduced by a factor 2-3 upon solvation, whereas the
reduction is of the order of a few tens percent only in the case of
light molecules; {\em ii)} the inertia of solvated heavy rotors (such
as, {\em e.g.}, OCS or SF$_6$) would reach the large-size
(nano-droplet) regime well before the first solvation shell is
completed, whereas the convergence for light rotors would be much
slower. This scenario was partially rationalized using the concept of
{\em adiabatic following}.  Heavy molecules rotate slowly when excited
and they usually have a stronger and more asymmetric interaction
potential with the surrounding He atoms. As a result of the
combination of these two effects, a fraction of the He solvent density
would be dragged along by the rotation of the molecule, thus
contributing to increase its inertia. This effect would be greatly
reduced in light rotors both because they rotate faster, and because
their interaction with He atoms is usually more spherical. Both these
facts hinder the dragging of He atoms, in qualitative agreement with
experiments which positively support the first of the above two
findings.  The fact that only a fraction of the He density in the
first solvation shell would adiabatically follow the molecular
rotation would explain the alleged fast convergence of the molecular
inertia to the nano-droplet regime for heavy rotors
\cite{focus_article,toennies}. The alleged slow convergence for light
rotors would instead be due to the interaction between molecular
rotation and bulk-like collective excitations that is supposed to
occur for large enough clusters \cite{focus_article,viel,zillich}.
Experimental results often exist only for large clusters. Some results
on small clusters begin to be available, but data in the intermediate
regime are not yet sufficient to validate this picture.
  
In this paper this problem is re-addressed using {\em Reptation
  Quantum Monte Carlo} (RQMC), a method that we believe to be the 
most appropriate existing to date to study the dynamical properties of
interacting boson systems \cite{NoAntri-Cornell,NoAntri-PRL}. As
paradigmatic cases of heavy and light solvated molecular rotors we
choose OCS@He$_N$ and HCN@He$_N$, respectively. In order to
disentangle the role of the bare molecular inertia from that of the
potential anisotropy, we have performed computer experiments in which
the rotational dynamics of OCS and HCN molecules was simulated using a
fictitious inertia appropriate to the {\em other} molecule. These
experiments indicate that the approach to the nano-droplet regime, as
well as the reduction of the molecular inertia upon solvation, is
determined by the anistropy of the potential, more than by the
molecular weight.

Our paper is organized as follows. In Sec. II we briefly review our
theoretical framework. Sec. III.A contains our results for OCS@He$_N$
which considerably extend the size range considered in a previous
publication \cite{NoAntri-2003}. We also address the problem of the
accuracy of the molecule-He interaction potential \cite{HH} and we
show that avoiding the {\em morphing} procedure followed to brush up
the potential, allows for a much better (actually, almost perfect)
agreement between simulations and rotational spectroscopy experiments.
In Sec. III.B we present results of similar simulations made for
HCN@He$_N$.  In Sec. III.C we present computer experiments done on
molecules with {\em fudged} masses ({\em i.e.} by using the inertia of
one molecule and the interaction potential of the other), and we
discuss the implications of these results on the main issue raised in
the present paper ({\em i.e.} the relative importance of the bare
molecular inertia and of the anisotropy of the molecule-atom potential
in determining the rotational properties of solvated molecules, as a
function of the cluster size). Finally, Sec.  IV contains our
conclusions.

\section{Theory}

\begin{figure}
  \begin{center}
    \includegraphics[width=8cm]{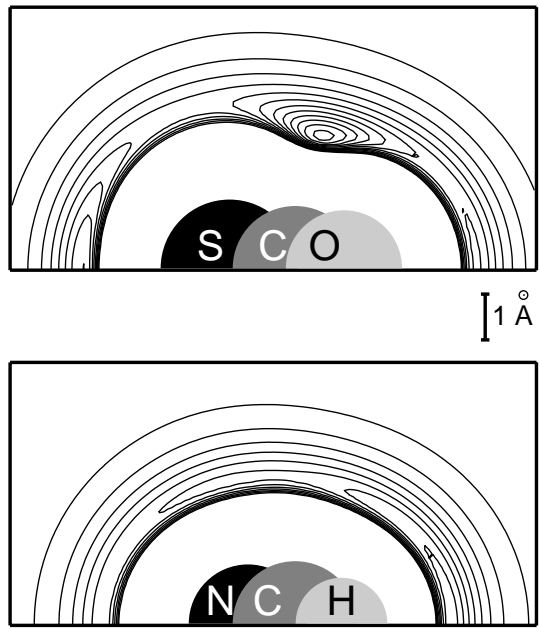}
  \end{center}
  \caption{Potential energy surfaces of He-OCS (up) and He-HCN (down)
    used in this work. The He-OCS PES is the {\em unmorphed} version
    from Ref.~\onlinecite{HH} (see text).  Contour levels start from
    $V=0$ to negative values, and they are spaced by $5~^{\circ}\rm
    K$.  The He-OCS interaction potential has a deep {\em equatorial}
    doughnut-shaped well around the C atom, and two secondary {\em
      polar} minima in correspondence of the O and S atoms, the one at
    the S pole being broader and deeper.  The He-HCN interaction
    potential, instead, has a shallow main minimum in correspondence
    of the hydrogen molecular pole and an even shallower secondary one
    slightly below the equator on the nitrogen side, the latter being
    barely visible in the figure.
    \label{fig:pes}
  }
\end{figure}

We consider a realistic Hamiltonian in which $N$ He atoms are treated
as point particles interacting via a pair potential, and the dopant
molecule (either OCS or HCN) as a rigid linear rotor, with only
translational and rotational degrees of freedom. Both the He-He and
the He-molecule interactions are parametrized after accurate ab-initio
quantum chemistry calculations \cite{HH,Korona,Cybul}. For the He-OCS
potential energy surface (PES), in particular, we use the {\em
  unmorphed} version of Ref.~\onlinecite{HH}. In the next section we
show that this version---which results from a direct fit to
quantum-chemistry calculations---is considerably more accurate for a
wide range of cluster sizes \cite{Genova} than the {\em morphed} one
which was refined so as to improve the predictions for the spectrum of
the OCS@He$_1$ complex \cite{HH}.  Contour plots of the He-HCN and the
He-OCS PES are displayed in Fig.~\ref{fig:pes}, showing a much greater
strength and anisotropy of the latter.

We simulate the system using the RQMC method
\cite{NoAntri-PRL,NoAntri-Cornell}. Details of the calculations are
similar to those of previous studies for OCS@He$_N$ and CO@He$_N$
clusters \cite{NoAntri-2003,NOI-CO}. Briefly, the imaginary-time
evolution operator projects a state defined by a trial function
$\Phi_0$ onto $|\Phi(\beta)\rangle=\exp(-\beta\widehat
H)|\Phi_0\rangle$, which in turn approaches the exact ground state,
$\Psi_0$, as $\beta$ goes to infinity. Expectation values on
$|\Phi(\beta) \rangle$, $\langle\Phi(\beta)|\widehat{\cal
  O}|\Phi(\beta)\rangle/ \langle\Phi(\beta)|\Phi(\beta)\rangle $, are
computed using a discretized path-integral representation of the
imaginary-time evolution, which becomes exact in the limit of zero
time step. The resulting paths are sampled with an efficient
generalized Metropolis algorithm. The average $\langle \widehat {\cal
  O}\rangle$ of the values taken by the operator $\widehat {\cal O}$
on the sampled paths gives an estimate of the desired expectation
value within a known statistical error and without any mixed-estimate
nor population-control biases. Here $\widehat{\cal O}$ can be either
a static operator (such as the Hamiltonian itself or the atomic number
density distribution) or a time-dependent one (such as those needed to
calculate time-correlation functions: $\widehat {\cal
  A}\exp(-\tau\widehat H)\widehat {\cal B}$. A single path thus
entails the imaginary-time evolution of the system for a total time
$2\beta+\tau$, which in our simulations is typically between 1 and 2
inverse K, with a discretization time step, $\epsilon$, of
10$^{-3}~{\rm K}^{-1}$. By explicitly checking the convergence of the
results to the limits of large projection time and zero time step, we
make sure that the computed quantities accurately represent their
ground state values.

The trial function is taken of the Jastrow form,
\begin{equation}
  \Phi_0 = \exp \left [ -\sum\limits_{i=1}^N u_1(r_i,\theta_i)
    -\sum\limits_{i<j}^N u_2(r_{ij}) \right ], 
\end{equation}
where ${\textbf r}_i$ is the position of the $i$-th atom with respect
to the center of mass of the molecule, $r_i=|{\textbf r}_i|$,
$\theta_i$ is the angle between the molecular axis and ${\textbf
  r}_i$, and $r_{ij}$ is the distance between the $i$-th and the
$j$-th He atoms.  On account of the anisotropy of the PES, the
He-molecule pseudo-potential, $u_1$, is expressed as a sum of products
of Legendre polynomials times radial functions. Retaining five or six
terms in the sum is enough to represent even the most anisotropic
situation, namely a ring of up to 5 He atoms tightly bound around the
OCS molecular axis in correspondence of the principal minimum of the
PES. The choice of the radial functions is not critical, as soon as
their variational parameters ensure sufficient flexibility in the
relevant range of He-molecule distances. The resulting variational
parameters, whose number is in the range of several tens, are
optimized by minimizing the variational energy with the variational
Monte Carlo method.

Information on the rotational excitations of the cluster can be
obtained from the imaginary time autocorrelation functions
\cite{NoAntri-2003}:
\begin{equation}
c_J(\tau) = \langle P_J\left({\bf n}(\tau)\cdot {\bf n}(0)\right)\rangle,
\label{c_of_t}
\end{equation}
where ${\bf n}$ is the molecular orientation versor, and $P_J$ is the
Legendre polynomial of degree $J$.  A generic imaginary-time
correlation function can be represented as a linear combination of
exponentials, whose decay constants are the excitation energies of the
system, and whose coefficients are the corresponding oscillator
strengths. The usefulness of this representation for obtaining dynamical
properties from quantum simulations is in general rather limited.
This is so because the calculation of a spectrum from imaginary-time
correlations entails carrying out an inverse Laplace transform, a
notoriously ill-conditioned problem \cite{maxent}. In the present
case, however, the situation is not as bad because very few excited
states contribute to $c_J(\tau)$. In fact, if the solvated molecule
were isolated, only one rotational state would contribute to
$c_J(\tau)$ which would therefore have the form of a single
exponential. Furthermore, the interaction between the solvent matrix
and the soluted molecule is rather weak, so that this
single-exponential picture is only slightly perturbed. Last, and most
important, the bosonic nature of the quantum solvent determines a low
density of low-energy excitations. As a consequence of the scarcity of
low-lying excitations available to couple with the molecular rotation,
it is thus possible to reliably extract their energies by a
multi-exponential fit.  A more general analysis, based on the maximum
entropy method, gives equivalent results \cite{paesani2004}.

\section{Results and discussion}
\subsection{Carbonyl sulfide}
Carbonyl sulfide (OCS) appears to have a special role in the
spectroscopy of He-solvated molecules. OCS was used as a microscopic
rotor to probe the superfluidity of He nano-droplets in a celebrated
{\em microscopic Andronikashvili experiment} \cite{toennies-OCS}. As a
consequence of the paucity of low-lying excitations in the bosonic
host, molecular rotational lines are exceedingly sharp
and well resolved when the molecule is solvated in $^4$He, while they
are broad and unresolved in $^3$He. In the former case the rotational
spectrum is well represented by a quasi-rigid free-rotor model:
\begin{equation}
  E_J = B J(J+1) - DJ^2(J+1)^2,
\end{equation}
where $B$ is the effective rotational constant of the
molecule---renormalized by the interaction with the host to a value
smaller but of the same order as in the gas phase---and $D$ is the
rotational distortion constant which is several orders of magnitudes
larger than for the free molecule. 

A few years later, size-resolved rotational and vibrational spectra of
small ($N=1\div 8$) OCS@He$_N$ clusters appeared \cite{jaeger-OCS},
indicating that fingerprints of superfluidity can be identified in
clusters as small as $N=8$. These findings spurred a burst of
theoretical research aimed at elucidating the relations existing among
structure, dynamics, and superfluidity in these small clusters
\cite{whaley-2003,NoAntri-2003}. A semi-quantitative agreement between
POITSE results from diffusion quantum Monte Carlo simulations and
experiments was reported in Ref. \onlinecite{whaley-2003} in the size
range $N=1\div 8$ (a minimum in the rotational constant was predicted
for $N=6$, while experiments show a decrease up to the largest
resolved size, $N=8$), while for $N>6$ a fast convergence to the
nano-droplet limit was observed (full convergence was claimed to be
achieved for $N=20$). A better agreement was found in Ref.
\onlinecite{NoAntri-2003}, in that the minimum of the rotational
constant (located at a cluster size $N=8$) was found to be compatible
with spectroscopic data. Results were also presented for the
rotational distortion constant, $D$, which was found to reach a
minimum for $N=5$, in agreement with experimental findings
\cite{jaeger-OCS}. In addition, a detailed analysis---based on a
comparison between the structure of the clusters and the He-OCS
angular-current correlations---allowed to shed light on the
microscopic mechanisms responsible for superfluidity.

\subsubsection{Appraising the quality of intermolecular potentials}

\begin{figure}
  \begin{center}
    \includegraphics[width=8cm]{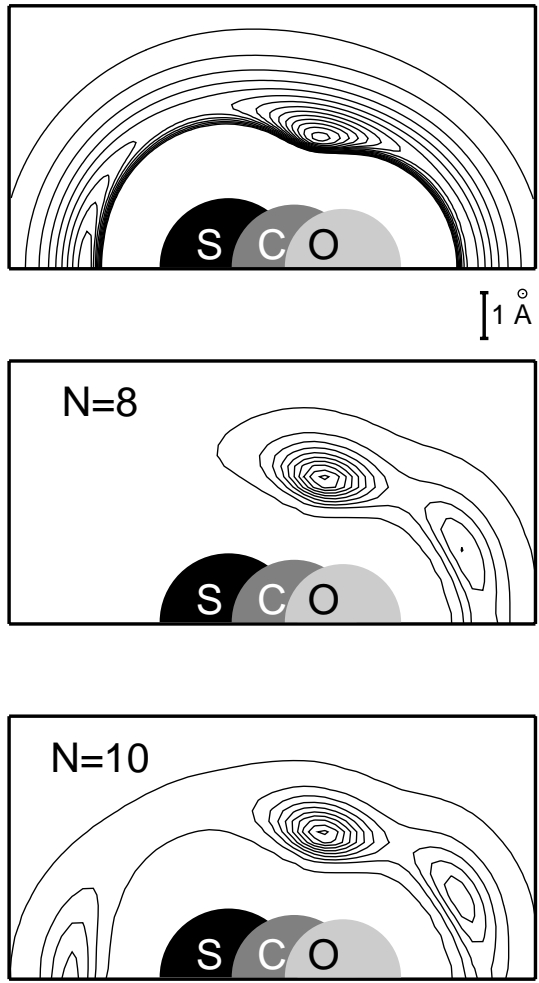}
  \end{center}
  \caption{
    Upper panel: potential energy surface of He-OCS (we display here
    the {\em morphed} version from Ref.~\onlinecite{HH} which was used
    in our previous study \cite{NoAntri-PRL}). Lower panels: contour
    plots of the helium density profiles of OCS@He$_8$ and
    OCS@He$_{10}$. For the density profiles, contour levels start from
    0.001 with increments of 0.005, in units of \AA$^{-3}$.
    \label{fig:OCS-He8}
  }
\end{figure}

Much of the discrepancy between the results of quantum Monte Carlo
simulations and experiments (as well as {\em among} different
simulations, we should add) may be due to the quality of the
He-molecule potential utilized for the simulations
\cite{NoAntri-2003}. In Ref. \onlinecite{paesani2004} it was in fact
shown that the difference between the predictions of Refs.
\onlinecite{NoAntri-2003} and \onlinecite{whaley-2003} is indeed
(almost) entirely due to the poorer quality of the potential utilized
in Ref.  \onlinecite{whaley-2003} \cite{footnote}. In Fig.
\ref{fig:OCS-He8} we compare the atomic density distributions of
OCS@He$_8$ and OCS@He$_{10}$ with the OCS-He potential used in Ref.
\onlinecite{NoAntri-2003}. According to the analysis of Ref.
\onlinecite{NoAntri-2003}, the minimum in the effective inertia of the
solvated OCS molecule occurs at the largest cluster size at which
quantum tunneling between the main, {\em equatorial}, and the two
secondary, {\em polar}, potential wells is hindered by the energy
barriers which separate them. It is clear that the larger the barrier,
the smaller the tunneling, and the larger the corresponding effective
inertia will be. In Ref.  \onlinecite{NoAntri-2003} it was indeed
found that fudging the OCS-He PES---so as to enhance the potential
energy barrier which separates the main well from the molecular
poles---hardly affects the rotational constant for $N\le 5$, while it
increases the inertia for $N=6,7,8$, thus bringing the results of the
simulations in much better agreement with experiments. It was later
found that this too low a value for the relevant energy barriers is in
fact an artifact of the {\em morphing} procedure adopted in Ref.
\onlinecite{HH}, so as to bring an already very accurate OCS-He
potential obtained from coupled-cluster quantum-chemical calculations
into an even better agreement with the known spectra of the OCS@He$_1$
complex. Such spectra depend on the details of the potential around
the minimum more than on the height of the barriers which separate
inequivalent minima. The latter, instead, are expected to crucially
affect the properties of the cluster when the main potentials well
starts to be filled by He atoms ($N\ge 5$, in the present case). We
conclude that the spectra of clusters with many He atoms---which
depend on the PES far from the minimum---are a much more sensible
benchmark of the quality of the PES than those of the helium-molecule
dimer. A similar conclusion was drawn in Ref.
\onlinecite{paesani2004} where a different He-OCS PES was proposed,
such that the rotational constants calculated from POITSE simulations
agreed well with experimental data available for $N\le 8$. RQMC
simulations performed with this PES in the size range $10 \alt N \alt
30$ show a behavior of the rotational constant very similar to that
reported in Fig. \ref{fig:rot-cnst-OCS}, obtained from the {\em
  unmorphed} version of the PES of Ref.  \onlinecite{HH}, although the
resulting values are slightly smaller
for the largest sizes
(hence closer to the experimental nano-droplet regime: see below). 

\subsubsection{Effective rotational constants}

\begin{figure}
  \begin{center}
    \includegraphics[width=8cm]{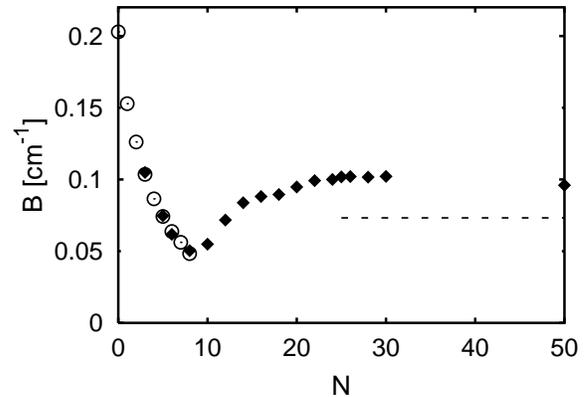}
  \end{center}
  \caption{
    Effective rotational constant of OCS@He$_N$, as a function of the
    cluster size, $N$. Full diamonds: results from RQMC simulations
    performed using the unmorphed potential of Ref. \onlinecite{HH}
    (see text). Open circles: experimental data from Ref.
    \onlinecite{jaeger-OCS}. The horizontal dashed line indicates the
    nano-droplet limit measured in Ref. \onlinecite{jaeger-OCS}.
    \label{fig:rot-cnst-OCS}
  }
\end{figure}

\begin{figure}
  \begin{center}
    \includegraphics[width=6.0cm,angle=270]{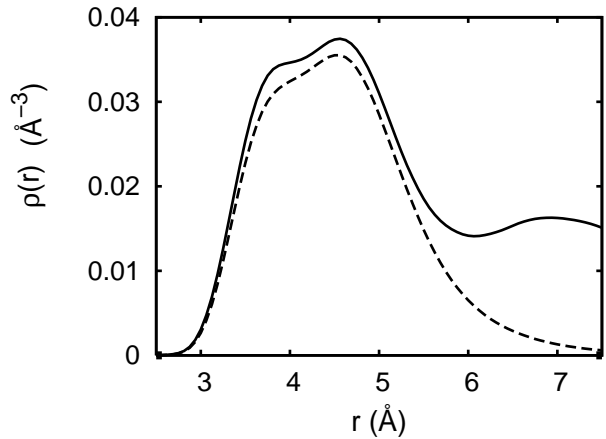}
  \end{center}
  \caption{
    Radial density profile $\rho(r)$ of He in OCS@He$_N$ for $N=20$
    (dashed line) and $N=50$ (solid line).
    \label{fig:rad-rho}
  }
\end{figure}

In Fig. \ref{fig:rot-cnst-OCS} we report the dependence of the
effective rotational constant, $B$, on the cluster size, $N$, as
calculated with the unmorphed He-OCS potential of Ref.
\onlinecite{HH}. The nano-droplet limit resulting from measurements on
clusters of $\approx 6000$ atoms \cite{toennies-OCS} is indicated by a
dashed line. Our results compare favorably with experimental data
which are available up to $N=8$ \cite{jaeger-OCS}. In the small-size
regime ($N\le 8$) the rotational constant decreases with increasing
cluster size because the He atoms trapped in the ({\em equatorial})
and in the first {\em polar} (near oxygen) well are dragged along by
the molecular rotation, thus increasing the effective inertia. For
$N>8$---as the second {\em polar} (near sulphur) well starts to be
filled---an increasing fraction of the He atoms can freely tunnel
among different wells, thus not contributing to the molecular inertia.
The quantum fluid nature of the He solvent is such that tunneling is a
collective process in which not only the excess atoms not fitting in
the main potential well take part, but even those which are tightly
bound to it. For this reason, once tunneling among different wells is
made possible by the spill-out of excess atoms, this process
determines a decrease of the molecular inertia, {\em i.e.} an increase
of the rotational constant for an increasing cluster size. As a matter
of fact, it was shown in Ref. \onlinecite{NoAntri-2003} that the value
of the correlation between the molecular angular momentum and the
atomic angular current---which is maximum in the main {\em equatorial}
potential well---starts decreasing as the rotational constant
increases past the minimum ($N>8$). This decrease continues until the
first solvation shell is completed at a cluster size $N\approx 20$,
around which the rotational constant seems to stabilize. As the second
solvation shell starts to build, however, quantum exchange cycles
involving atoms from this shell would contribute to further, although
weakly, decrease the molecular inertia.  A similar behavior of the
evolution of the rotational constant for sizes shortly beyond
completion of the first shell is also observed in simulations of
clusters doped with N$_2$O and CO$_2$
\cite{moroni-roy,moroni-mckellar}, which are all molecules having a
qualitatively similar interaction with He atoms.  Our findings
demonstrate that---contrary to a commonly accepted assumption---He
atoms from outer (larger than the first) solvation shells do affect
the molecular inertia. For N$_2$O and CO$_2$ this is also supported by
experimental evidence \cite{jaeger-N2O,mckellar-CO2}: the measured
value of $B$ for the largest cluster with secure assignment of
spectral lines ($N=12$ for N$_2$O and $N=17$ for CO$_2$) is
significantly higher than the nano-droplet limit, with no plausible
signs of convergence within the first shell.  A role of outer shells
in bringing down the effective rotational constant is thus to be
expected. How far from the molecule does this effect extend, our
simulations---which are limited at present to a few tens
atoms---cannot say yet. This finding is at least compatible with
current phenomenological models of the inertia of He-solvated
molecules, which predict a lower contribution of the solvent to the
molecular inertia, with decreasing atomic density. In hydro-dynamical
models a lower atomic density would determine a reduced kinetic energy
of the irrotational flow of the solvent; in a two-fluid model,
instead, a lower inertia would simply follow from a reduction of the
non-superfluid component of the atomic density. Coming down from the
nano-droplet regime, a decrease of the atomic density in the first
solvation shell(s) is indeed expected, as a consequence of the reduced
pressure exerted by the outer shells.  This is clearly demonstrated in
fig.~\ref{fig:rad-rho} in which the He radial density profiles around
OCS are compared for $N=20$ and $N=50$.  The radial density in the
first shell is significantly lower for the smaller cluster.  On purely
classical grounds, a competing effect could arise if the spatial
extension of the first shell was sensitive to the pressure release,
thus affecting the second moments of the atomic density. The density
profiles shown in Fig.~\ref{fig:rad-rho} however suggest that this
effect is small, since the positions of the peaks in the first shell
hardly change between clusters of 20 and 50 He atoms.

Our results demonstrate that in OCS@He$_N$ clusters the effective
rotational constant does not attain its asymptotic limit upon
completion of the first solvation shell, being in fact higher at this
size. When the second shell starts to build up, the value of $B$
further increases (arguably, via quantum exchange cycles involving
particles of both shells \cite{draeger,moroni-roy}). Whether, in
larger clusters, the missing inertia will be recovered by a change in
the density of the first shell or by a direct contribution from the
outer shells, or both, remains to be investigated.

\subsection{Hydrogen cyanide}

Hydrogen cyanide (HCN) is considered as a prototype of light
helium-solvated rotors \cite{toennies,focus_article}, for which the
assumption of {\em adiabatic following} breaks down. This has been
demonstrated both experimentally, by comparing the rotational
constants of HCN and DCN in the nano-droplet regime \cite{adiab-Sc},
and theoretically, by comparing the He density profiles obtained from
simulations performed including and, in turn, neglecting the molecular
degrees of freedom \cite{viel}. The calculation of the molecular
rotational constant, B, in large droplets thus defies the application
of either the two-fluid \cite{focus_article} and hydrodynamic models
\cite{callegari}, which both rely on {\em adiabatic following}, albeit
in a different manner. On the other hand, direct calculation of
rotational excitations by the POITSE method \cite{viel} indicated that
the nano-droplet value would not be reached even for $N=25$, well
beyond completion of the first solvation shell. This slow convergence
of the rotational constant as a function of the cluster size was later
attributed to the coupling of molecular rotation to phonon-like
excitations of the solvent which would develop only in the
nano-droplet regime \cite{zillich}. Unfortunately experimental data
for the effective rotational constant of HCN are only available, so
far, for very large droplets \cite{adiab-Sc}. We will see however how
a comparison of our theoretical results for HCH@He$_N$ with both
theoretical and experimental results which are available for the
closely related CO@He$_N$ system will allow to draw a number of
important and non trivial conclusions.

\begin{figure}
  \begin{center}
    \includegraphics[width=8cm]{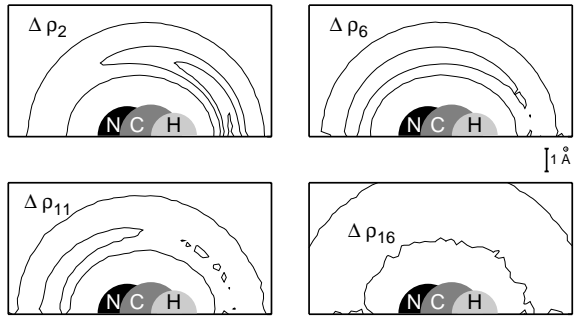}
  \end{center}
  \caption{
    Contour plot of the incremental density, 
    $\Delta\rho_N({\bf r})=\rho_N({\bf r})-\rho_{N-1}({\bf r})$, 
    for selected HCN@He$_N$ clusters.
    The HCN molecule has its center of mass at the origin,
    with the hydrogen atom on the positive $x$ direction.
    Distances are in $\rm\AA$, and contour levels start from 0.0001 with
    increments of 0.001, in units of inverse cubic $\rm\AA$.
    \label{fig:drho}
  }
\end{figure}

\subsubsection{Structural properties}

In Fig. \ref{fig:drho} we report the incremental atomic density
profile of HCN, $\Delta\rho_N({\bf r})=\rho_N({\bf r})-\rho_{N-1}({\bf
  r})$, as calculated for a few selected cluster sizes. For the binary
complex, HCN@He$_1$, the He density forms a broad cap around the H end
of the HCN molecule, centered on the main well of the PES. As $N$
increases, the He density smoothly piles up with no visible signature
of the secondary, {\em sub-equatorial}, minimum. The incremental
density is mainly localized on the H side of the HCN molecule for $N$
up to 6, mainly on the N side for $N$ between 8 and 14, and nearly
isotropically thereafter.  Starting from $N$ around 16, the
incremental density shifts toward larger distances from the HCN center
of mass (see the inner and the outer contour levels in
Fig.~\ref{fig:drho}). We consider this behavior as an indication that
the first solvation shell is completed around $N\approx 15$, although
shell effects are by no means sharp in this system (for instance, much
more pronounced effects are seen at $N=12$ in clusters of
para-hydrogen seeded with HCN \cite{botti,melting}).

\begin{figure}
  \begin{center}
    \includegraphics[width=8cm]{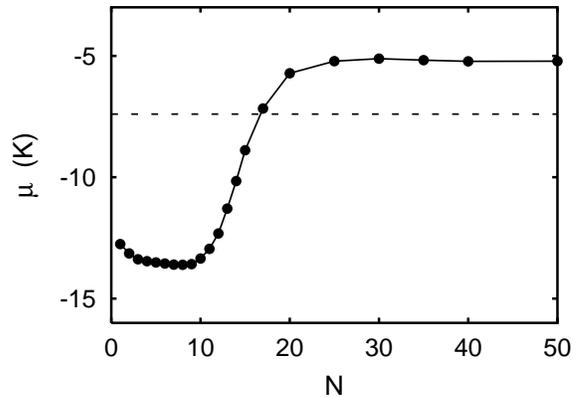}
  \end{center}
  \caption{
    Chemical potential, $\mu(N)=E(N)-E(N-1)$, for HCN@He$_N$ clusters.
    The dashed line is the bulk limit.
    \label{fig:mu}
  }
\end{figure}

The dependence of the cluster ground-state energy, $E(N)$, on the size
of the system reflects the extremely smooth evolution of the He
density profile, as shown in Fig.~\ref{fig:mu} which reports the
chemical potential, $\mu(N)=E(N)-E(N-1)$. For $N$ up to a dozen the
effects of the He-HCN interaction dominates: the chemical potential
stays almost constant, 
decreasing only very weakly with the
cluster size, as a consequence of the He-He interaction. For
$12 \alt N \alt 20$ the first solvation shell is filled
and the kinetic energy of He atoms increases due to their closer
packing, thus determining a 
rise of the chemical potential. For
$N\agt 20$ $\mu(N)$ is stabilized again at value which is 
higher
than the bulk limit, due to the smaller effects of the He-He
interaction in this size range.

\begin{figure}[h]
  \begin{center}
    \includegraphics[width=8cm]{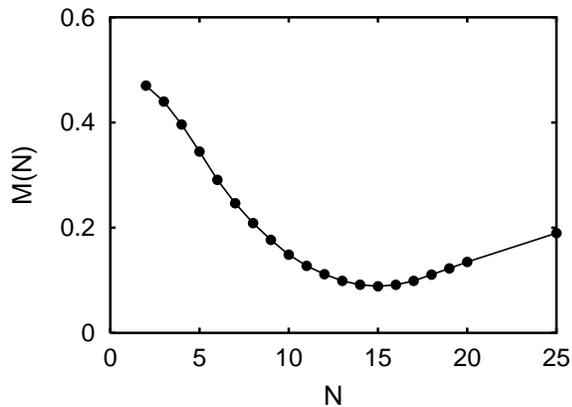}
  \end{center}
  \caption{
    Integrated pair distribution of the dihedral angle,
    defined in Eq.~\ref{eq:M}, as a function of the cluster size,
    for HCN@He$_N$.
    \label{fig:M}
  }
\end{figure}

The fact that the first solvation shell is completed in the size range
$12 \alt N \alt 20$ is confirmed by an analysis of the
angular correlations of He atom pairs. Let us define
the He-He pair distribution of the dihedral angle with respect to the
molecular axis, $\phi$, as:
\begin{equation}
  C(\phi)={1\over N(N-1)} \left \langle\sum_{i<j}\delta(\phi-\phi_i+\phi_j)
  \right \rangle.
\end{equation}
$C(\phi)$ gives a measure of the (weak) tendency of the He atoms to cluster
together on the same side of the molecular axis, due to the He-He
attraction \cite{NOI-CO}. The integrated
quantity, 
\begin{equation}
  M=\int_0^{\pi/2}C(\phi)d\phi-1/2, 
  \label{eq:M}
\end{equation}
displayed in Fig.~\ref{fig:M} as a function of $N$, reveals that this
tendency is smallest around $N=15$, corresponding to completion of the
first solvation shell where, for steric reasons, the angular
distribution of He atoms around the molecule is most uniform.

\begin{figure}[h]
  \begin{center}
    \includegraphics[width=8cm]{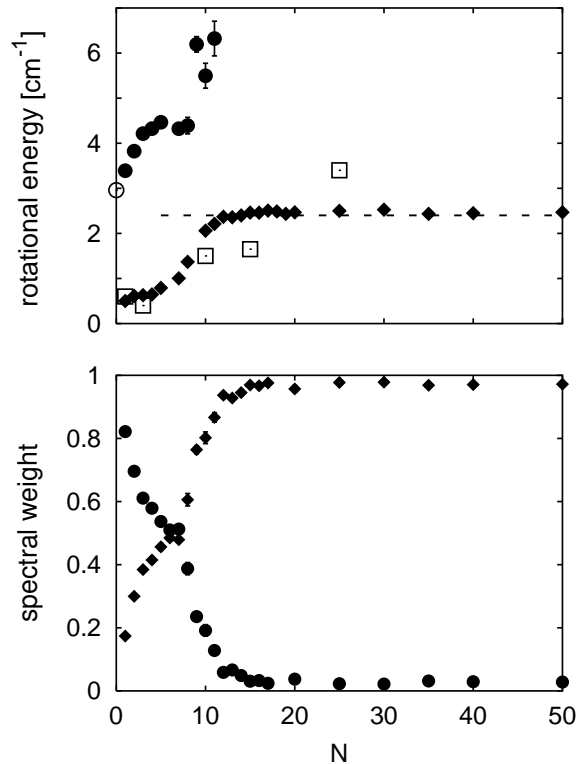}
  \end{center}
  \caption{
    Upper panel: Rotational energies of HCN@He$_N$ as functions of the
    cluster size, computed by RQMC ($a$-type, diamonds and $b$-type,
    filled circles);
    results from the POITSE calculations of Ref.~\onlinecite{viel} are
    indicated by squares.
    The dashed line is the effective rotational constant of HCN in the
    nano-droplet limit\cite{adiab-Sc}, while the empty circle at $N=0$ is
    the gas-phase value. Energy units are cm$^{-1}$. Lower panel: Spectral
    weight of the $a$-type line (diamonds) and the $b$-type line
    (filled circles).
    \label{fig:rot-e}
  }
\end{figure}

\subsubsection{Rotational excitations}

All the above features of He-solvated HCN molecules are similar to the
CO case, already studied in a previous work \cite{NOI-CO}. In
Fig.~\ref{fig:rot-e} we report the energies of the two lowest
rotational excitations with angular momentum $J=1$, along with the
corresponding spectral weights. We note that our predictions for
$N>10$ considerably differ from those of Ref. \onlinecite{viel} which
were obtained with the POITSE method. We cannot offer any explanation
for this discrepancy. We can only observe that the value of $B$
calculated in Ref. \onlinecite{viel} for $N=25$ is even larger than in
the gas phase, a fact that can hardly be explained on physical
grounds. In fact, while the reduction of one of the moments of inertia
below its value in gas phase---experimentally observed in the binary
complex---can be explained by the considerable relative mass
redistribution occurring upon formation of the binary complex, similar
effects are hard to justify at cluster sizes where the density profile
evolves with $N$ in a smooth and almost isotropical way. This seems to
suggests that the results of Ref. \onlinecite{viel} may be affected by
some inaccuracies for the largest sizes considered in that work. On
the other hand, while it is difficult, for large cluster sizes, to
ensure full convergence with respect to projection time, as well as
full ergodicity in path sampling, we note that the present
calculations for HCN are indirectly supported by the agreement between
the experiment \cite{mck-co} and similar calculations for the closely
related CO-He clusters \cite{NOI-CO}.

Two series of excitations, called {\em $a$}-type and {\em $b$}-type
lines, evolve smoothly from the known {\em end-over-end} and the {\em
  free molecule} rotational modes of the binary complex, respectively.
The $b$-type line starts off with a stronger spectral weight, then it
quickly weakens and eventually disappears. Note how the intensity of
the $b$-type line follows the decline of the He-density anisotropy
displayed in Fig. \ref{fig:M}. The existence of two relevant spectral
lines (instead of a single line appropriate to the spectrum of a
linear rotor) is in fact a manifestation of the dynamical anisotropy
of the atomic-density distribution around the molecular axis, which
lowers the cylindrical symmetry of the molecular rotor. For $N>15$ the
rise of the He-density anisotropy does not give rise to any
significant line splitting because it is due to atoms in the second
solvation shell, which are only very weakly coupled to the molecular
rotation.

\begin{figure}
  \begin{center}
    \includegraphics[width=8cm]{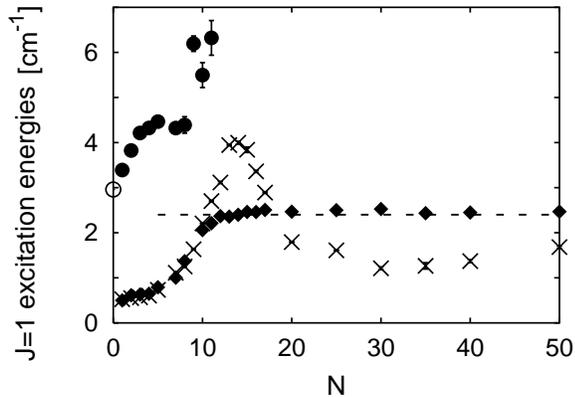}
  \end{center}
  \caption{
    Comparison between the positions of the spectral lines reported in
    Fig. \ref{fig:rot-e} with the lowest mode in the spectral
    resolution of the correlation $\gamma(t)$.
    \label{fig:gamma-e}
  }
\end{figure}

The coupling between the molecular rotation and He-density
fluctuations is best understood using the (imaginary) time
correlations of the versor, $\bf u$, pointing from the molecular
center of mass towards the center of mass of the complex of He atoms.
Such a time correlation function, $\gamma(t)=\langle {\bf u}(0)\cdot
{\bf u}(t)\rangle$, contains information on the energies and spectral
weights of the $J=1$ cluster excitations whose character is
predominantly that of a He-density fluctuation. In Fig.
\ref{fig:gamma-e} we display the lowest-lying excitation energy
extracted from the spectral resolution of $\gamma(t)$ as a function of
the cluster size, and compare it with the positions of the {\em
  a-type} and {\em b-type} lines already reported in Fig.
\ref{fig:rot-e}. We see that this density-fluctuation
excitation---which has a relatively strong spectral weight---is
degenerate with the $a$-type line for $N\alt 10$. This mode has been
interpreted as a cluster excitation in which part of the He density is
dragged along by the molecular rotation \cite{NOI-CO}. This
interpretation is trivial for the binary complex (it corresponds to
the end-over-end rotation), but it has no obvious visualization for
larger clusters.  For $N>10$, the energy of this mode departs from the
$a$-type line, and the $b$-type line correspondingly disappears. This
relates to the onset of a situation where the He density is decoupled
from the molecular rotation.

We come now to the main concern of this paper, {\em i.e.}  the
convergence of the effective rotational constant to its asymptotic,
large-size, value. Our results indicate that the residual
renormalization of the effective inertia does not change significantly
upon further growth of the cluster beyond, say, $N=15$. This fact is
in complete analogy with the findings reported in
Ref.~\onlinecite{NOI-CO} for clusters doped with CO (and in agreement
high resolution IR spectra recently obtained for He clusters seeded
with CO up to $N=20$ \cite{mck-co}), but at variance with the results
reported for HCN in Ref. \onlinecite{viel} and with a commonly
accepted view. With a similar rotational constant and a similar
interaction with He, CO and HCN are not expected to behave very
differently upon solvation in He clusters. The experimental results
for CO do not seem compatible with a large variation of $B$ between,
say, $N=15$ and 25, thus challenging the idea of slow convergence to
the nano-droplet limit as a general feature of the effective inertia of
quantum solvated light rotors. Although a conclusive answer will
require the measurement and assignment of spectral lines for even
larger clusters, we believe that the agreement between our previous
calculations and high-resolution IR measurements for the closely
related CO@He$_N$ system warrants a considerable trust in the present
results and in the conclusions on the approach to the nano-droplet
regime based on them.

In conclusion, the results reported here for HCN and in
Refs.~\onlinecite{NOI-CO} for CO (the latter being supported by the
experimental study of Ref.~\onlinecite{mck-co}) suggest that for these
{\em light rotors} the asymptotic value of the effective rotational
constant is reached well before completion of the first solvation
shell. We will see in the next Section that this behavior is better
attributed to the weak anisotropy of the potential, rather than to the
small value of $B$ in gas phase.

\subsection{Fudged molecules}

In this section we consider the fractional reduction of the gas-phase
rotational constant $B_0$ upon solvation in He nano-droplets,
$\Delta=B/B_0$. The observed general trend (see e.g. Fig.~13 in
Ref.~\onlinecite{toennies}) is that lighter rotors tend to have larger
values of $\Delta$. A suitably defined {\em amount of adiabatic
  following} \cite{adiabatic_following} has been proposed as the key
physical property responsible for the value of $\Delta$.
Qualitatively, the analysis of Ref.~\onlinecite{adiabatic_following}
supports the simple picture that both a small molecular inertia and a
weakly anisotropic interaction lead to a large value for $\Delta$.
However, there are several exceptions: for instance, N$_2$O has
$B_0=0.30$~K and $\Delta=0.17$, whereas OCS (a {\em typical entry} in
the mentioned general trend) has $B_0=0.15$~K and $\Delta=0.36$. Given
that for OCS there is {\em saturation} to almost full {\em adiabatic
following}, \cite{adiabatic_following} the significantly smaller
$\Delta$ value for N$_2$O cannot be explained by resorting to the
concept of {\em adiabatic following} alone. Therefore, it seems useful to
gain further insight by disentangling the role of the PES anisotropy
from that of the gas--phase inertia. To this purpose, we have
performed simulations with two fictitious molecules: f-OCS ({\em
  fudged} OCS), with the PES as OCS and the same $B_0$ value as HCN,
and f-HCN ({\em fudged} HCN), featuring the HCN--He interaction and
the $B_0$ value of OCS. Note that the gas-phase rotational constants
of OCS and HCN are in a ratio of about 1:7.

\begin{figure}
  \begin{center}
    \includegraphics[width=8cm]{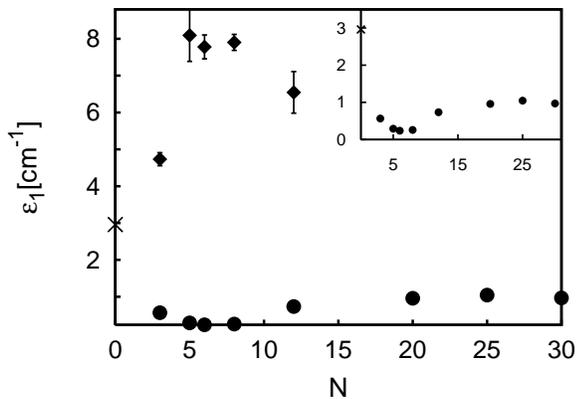}
  \end{center}
  \caption{
    Rotational energies of f-OCS@He$_N$ as a function of the
    cluster size ($a$-type, filled circles, and $b$-type, diamonds).
    The cross at $N=0$ shows the fictitious value of 2$B_0$.
    Inset: the detail of the $a$-type line.
    \label{fig:fakeocs-rot}
  }
\end{figure}

\begin{figure}
  \begin{center}
    \includegraphics[width=8cm]{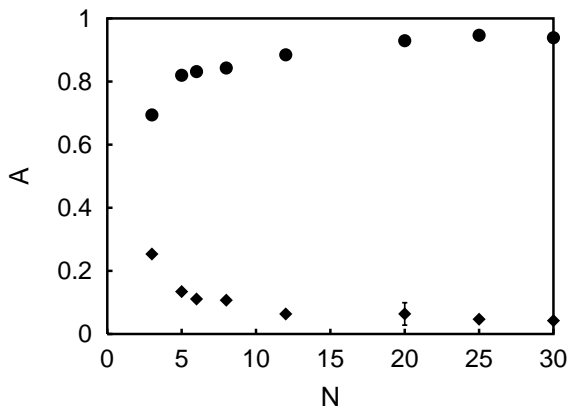}
  \end{center}
  \caption{
    Spectral weight of the $a$-type line (filled circles)
    and the $b$-type line (diamonds) for f-OCS@He$_N$.
    \label{fig:fakeocs-a}
  }
\end{figure}

The rotational energies of f-OCS@He$_N$ are shown in
Fig.~\ref{fig:fakeocs-rot}, and the corresponding spectral weights in
Fig.~\ref{fig:fakeocs-a}. The general appearence of both quantities
for f-OCS is closer to HCN than OCS, due to the presence of an
$a$-type and a $b$-type line, with the spectral weight of the latter
decaying with increasing $N$. However, at variance with HCN, the
$a$-type line for large $N$ approaches an energy significantly smaller
than the gas-phase value, $2B_0$. Taking the value at $N=30$ as an
estimate (very likely an overestimate, see Section III.A) of the
asymptotic value, we obtain $\Delta\simeq 0.33$, which is close
to---and somewhat smaller than---the value of OCS.

\begin{figure}
  \begin{center}
    \includegraphics[width=8cm]{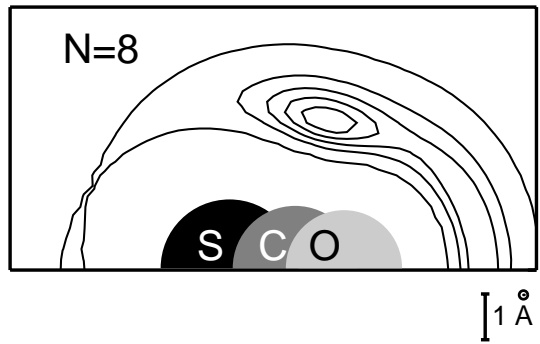}
  \end{center}
  \caption{
    Contour plot of the helium density profile of f-OCS@He$_8$.
    \label{fig:fakeocs-rho8}
  }
\end{figure}

The $a$-type line of f-OCS shows a minimum at $N=6$ or $7$, i.e. at a
smaller size than OCS. As shown in Fig.~\ref{fig:fakeocs-rho8} already
for $N=8$ there is a significant He density all around the molecule,
which is presumably responsible for the turnaround of the $B$ value
(although the relation between density profiles and turnaround of the
$B$ value could be not so straightforward for f-OCS, due to the
residual spectral weight in the $b$-type line). Regardless of the
implications for the turnaround, we stress that the significant
difference between the density profiles between f-OCS and OCS (see
Fig.~\ref{fig:fakeocs-rho8} and Fig.~\ref{fig:OCS-He8} for $N=8$)
implies a significantly different {\em amount of adiabatic following}. The
fact that the $\Delta$ values of f-OCS and OCS are nevertheless very
similar, indicates that the anisotropy of the potential, rather than
the dynamical regime implied by the gas-phase inertia, is mainly
responsible for the renormalization of the rotational constant upon
solvation.

\begin{figure}
  \begin{center}
    \includegraphics[width=8cm]{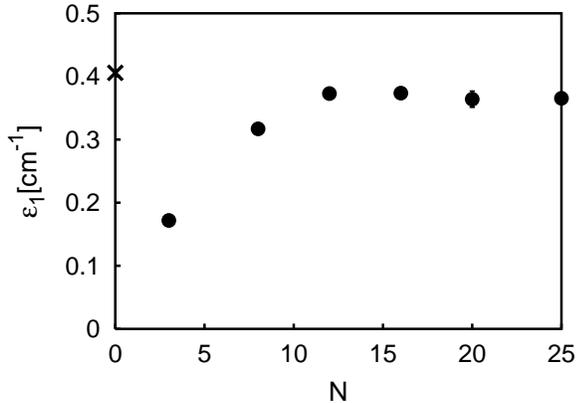}
  \end{center}
  \caption{
    Rotational energies of f-HCN@He$_N$ as a function of the cluster
    size. The cross at $N=0$ shows the fictitious value of 2$B_0$.
    \label{fig:fakehcn-rot}
  }
\end{figure}

Similar conclusions hold for f-HCN as well.  The energy of the lowest
rotational excitation with $J=1$ is shown in
Fig.~\ref{fig:fakehcn-rot} as a function of $N$. Already for $N=3$ the
spectral weight of this excitation exceeds 90 percent, and in this
respect f-HCN is closer to OCS than to HCN (i.e. the spectrum looks
that of a linear rotor for $N\ge 3$).  However as far as the value of
$\Delta$ is concerned, the effect of fudging the gas-phase inertia is
very small.

In order to estimate the asymptotic limit of the rotational constant,
we assume that it is given by the value at the largest-size cluster
simulated ($N=25$), noting that the evolution of $B(N)$ in
Fig.~\ref{fig:fakehcn-rot} is nearly flat for $N\geq 12$ (for HCN,
this assumption would give a very good agreement with the experimental
nano-droplet value, see Fig.~\ref{fig:rot-e}). This gives
$\Delta=0.90$ for f-HCN, close to---and somewhat higher than---the
value 0.81 measured in HCN \cite{adiab-Sc}.

Our findings indicate that the potential alone has a dominant role in
establishing the value of $\Delta$, at least in the range of physical
parameters appropriate to the linear molecules studied here.
Furthermore, the effect of reducing the molecular inertia while
keeping the PES fixed is to slightly {\em decrease} the $\Delta$
value. This result, in agreement with an experimental study of HCN and
DCN \cite{adiab-Sc}, contradicts the empirically established
correspondence between light rotational inertia and large $\Delta$
values. We conclude that such a correspondence is due, to a large
extent, to the generally small anisotropy of the interaction between
light rotors and Helium---an accidental effect, as far as the
rotational dynamics of the solvated molecule is concerned.
Within this picture, the behavior of N$_2$O is not an {\em anomalous}
case, but merely a consequence of the stronger stiffness and
anisotropy of the N$_2$O-He potential with respect to, say, OCS or
CO$_2$.

We finally note that a previous calculation for SF$_6$ with a
fictitiously small gas-phase inertia would support the opposite
conclusion that, for given interaction with the solvent, a lighter
molecule would have a larger value of $\Delta$ \cite{farrelly}. While
this result could shed some doubts on the generality of the
conclusions drawn from the analysis of a few {\em linear} molecules,
we believe that the calculations of Ref. \onlinecite{farrelly}
probably deserve further analysis because they were obtained using a
fixed-node approximation, whose accuracy is not warranted especially
for light rotors \cite{viel}.

\section{Conclusions}

Quantum Monte Carlo simulations, in conjunction with high-quality
inter-particle potentials, have reached a remarkable degree of
accuracy for the calculation of the rotational dynamics of molecules
solvated in He.  Using the RQMC method, we have studied the evolution
of the rotational excitations with the number $N$ of solvent He atoms
for a prototype heavy rotor, OCS, and a prototype light rotor, HCN
(for OCS, preliminary results have already been presented in
Ref.~\onlinecite{Genova}). The size range explored, larger than
presently attained with number-selective IR and/or MW spectroscopy,
allows us to draw a series of conclusions on the approach of the
rotational constant to its asymptotic value in the nano-droplet limit.
Our results entail a substantial revision, both quantitative and
qualitative, of the common view that the asymptotic limit would be
essentially determined by the amount of adiabatic following and
that---at least for heavy rotors---it would be reached well before
completion of the first solvation shell.

The rotational constant of OCS, after the undershoot of the
nano-droplet value and the turnaround which signals the onset of
superfluidity, crosses again its asymptotic limit at $N=12$; moreover,
starting with the beginning of the second solvation shell around $N
\approx 20$, it develops a further structure with an extremely broad
maximum, and a possible hint of a (final?) decrease only seen at the
largest size we studied, $N=50$.  This feature, which parallels
similar findings for CO$_2$ and N$_2$O, definitely supports {\em slow}
convergence to the nano-droplet value.  A strikingly different
behavior is found for HCN. In this case, a linear-rotor-like spectrum
is found for $N$ larger than 10, and the resulting rotational constant
stays constant in a wide range (say 15 to 50), with a value close to
the measured value in the nano-droplet limit. This result contradicts
the expectation of a slow convergence to the asymptotic limit
(determined by coupling of molecular rotation with well-developed
bulk-like excitations of the solvent) as a general property of light
rotors.  Indeed, a behavior very similar to that illustrated here for
HCN was found for another light rotor, namely CO.  It would be
tempting to propose that {\em fast} convergence to the nano-droplet limit
is a general rule for light rotors. However, preliminary experimental
results \cite{havenith} seem to indicate that, for CO, the asymptotic
limit is significantly lower than inferred from the nearly constant
value of the rotational constant in the size range from a dozen to a
few tens He atoms. The possibility of defining a general trend for
light rotors thus deserves further investigation.

In order to establish the relative importance of the bare molecular
inertia and of the strength and anisotropy of the He-molecule
interaction in determining the approach of the rotational dynamics to
the nano-droplet regime, we have also performed computer experiments
in which the molecular inertia was intentionally modified. To this
end, we have considered two {\em fudged} molecular species, f-OCS and
f-HCN, {\em i.e.} OCS and HCN with fictitiously small and fictitiously
large values of the gas-phase rotational constant, respectively
(appropriate in fact to the {\em other} molecule). Perhaps the most
important feature which was attributed to the predominant role of the
bare molecular inertia is the amount of renormalization of the
gas-phase rotational constant, $B$, upon solvation: the strong (weak)
reduction of $B$ observed for heavy (light) rotors was attributed to
the large (small) amount of adiabatic following. Our results indicate
that the fractional reduction of the gas-phase rotational constant
upon solvation is slightly {\em stronger} for f-OCS than for true OCS,
despite the obvious fact that adiabatic following is much larger for
the latter. Likewise, the reduction calculated for f-HCN is somewhat
{\em weaker} for f-HCN than for true HCN. The same trend was
experimentally observed, with a smaller variation of $B_0$, in a
comparative study of HCN and DCN \cite{adiab-Sc}. This clearly shows
that it is the strength and anisotropy of the He-molecule interaction,
rather than the bare molecular inertia, which is mainly responsible
for the renormalization of the rotational constant in the nano-droplet
regime. In this perspective, the classification into heavy and light
rotors thus retains its validity only to the extent that heavier
molecules tend to have stronger and more anisotropic interactions with
He.

\acknowledgments

We thank Giacinto Scoles for many fruitful discussions. Part of the
calculations presented in this work were done thanks to the {\em
  Iniziativa Calcolo Parallelo} of the Italian Institute for the
Physics of Matter (INFM).


\begin{thebibliography}{99}
\bibitem{footnote} 
  We have verified that the residual discrepancy visible in Fig. 6 of
  Ref.  \onlinecite{paesani2004} can be removed by improving the
  convergence of the RQMC simulations with respect to the length of
  the quantum paths ({\em snakes}).
\end{thebibliography}
\end{document}